\documentclass[%
,showpacs%
,amssymb, amsmath,nobibnotes, aps, prl]{revtex4}

\usepackage{graphicx}   

\begin{document}
	\title{
A possible demonstration of cooperative dynamical Jahn-Teller \\
effect in Raman scattering spectrum of rhombohedral $LaMnO_3$ 
}
\author{A.~E. Nikiforov}%
\author{S.~E. Popov}%
\email{Sergey.Popov@usu.ru}
\affiliation{Ural State University, pr. Lenina 51, Ekaterinburg, 620083 Russia}
\begin{abstract}
The Raman scattering (RS) spectra of rhombohedral $LaMnO_3$ is considered. It is shown that broad high intensity RS line at 640 $cm^{-1}$ \cite{Iliev1} could be explained in the framework of cooperative dynamical Jahn-Teller effect.  This hypothesis is illustrated through the lattice dynamics calculations explicitly allowing for the electron-phonon term both in crystal energy and in dynamical matrix. The averaging scheme for the one mode cooperative dynamical effect is proposed
\end{abstract}
\pacs{63.20.-e, 63.20.Dj, 78.30.Hv, 71.70.Ej}
        \maketitle
\section{Introduction} \label{SecIntroduction}
The colossal magnetoresistance (CMR) is the main feature of lanthanum manganites \cite{CMR1}. Because of metal to insulator transition, the doped rhombohedral manganites are of grate interest. 
The nature of ground state in manganites is still under the question. There are strong coupling between orbital (electronic) and lattice degrees of freedom in manganites due to Jahn-Teller (JT) effect on $Mn^{3+}$ ions. Such coupling gives rise to the formation of vibronic ground state \cite{BersukerPolinger}. The vibronic ground state fundamentally transforms magnetic and transport properties of manganites \cite{Millis1, ToyModel, Goodenough98}.

The pure $LaMnO_3$ is the model compound for investigating cooperative JT effect. Undoped $LaMnO_3$ crystallizes in two phases: orthorhombic and rhombohedral \cite{Andersen95, HuangSantoro}. The orthorhombic one features by the static cooperative JT effect. The static cooperative JT effect and $R\overline{3}c$ space symmetry of rhombohedral phase are incompatible because the JT distortions break this symmetry.
%
%
The only way to conserve the $R\overline{3}c$ space symmetry in rhombohedral phase is dynamical JT effect. The possibility of dynamical JT effect in manganites is widely discussed \cite{Millis1, Goodenough61, Goodenough98}. The dynamical JT effect manifests in two basic forms: uncorrelated local dynamical JT distortions of $[MnO_6]$ complexes and cooperative dynamical JT distortions, which can be called cooperative dynamical JT effect \cite{Sobolewski}.
The most powerful tool that could determine presence of low symmetry structure is Raman scattering (RS). We suppose that rhombohedral $LaMnO_3$ exhibits cooperative dynamical JT effect. 
We show that anomalous intensity of RS lines observed in rhombohedral $LaMnO_3$ \cite{Iliev1,MartinCarron02} could be explained in terms of cooperative dynamical JT effect. For the lattice dynamics calculation the interionic pair potential shell model approximation with electron-phonon (JT) term, which explicitly allows for the dynamical effect, is used \cite{NikiforovShashkin, NikiforovPopov00}.
\section{Symmetry and Qualitative Analysis of rhombohedral $LaMnO_3$}
As was mentioned by Iliev and coworkers \cite{Iliev1,Iliev2} the Raman spectra of rhombohedral $LaMnO_3$ differs strongly from RS of non JT $LaAlO_3$

\begin{figure}
\includegraphics[scale=1.5]{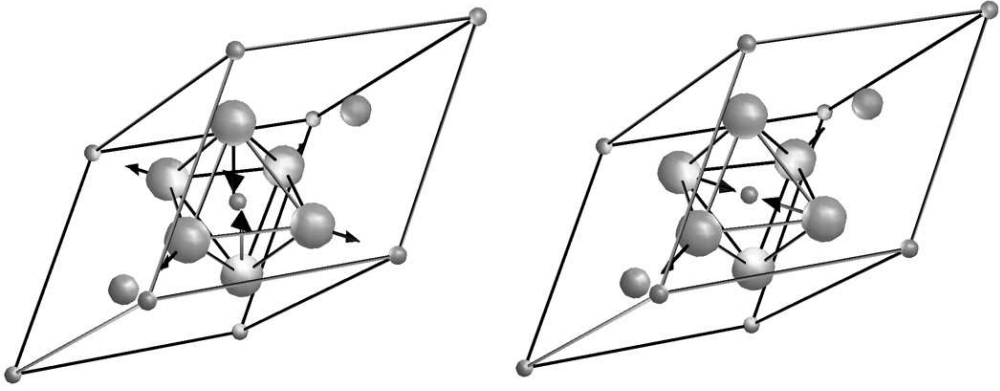}
\caption{\label{FigEg} The components of $E_g$ vibrational mode (left - $q_\theta$, right - $q_\varepsilon$)}
\end{figure}

From the symmetry analysis we know that in the $R\overline{3}c$ space group only five modes are active in RS ($A_{1g}+4 E_{g}$). The experiment on $LaAlO_3$ \cite{Iliev1} clearly shows 4 Raman active mode and one, that has very low frequency of $33\; cm^{-1}$ was also observed \cite{Scott}.
Comparing the RS of rhombohedral $LaMnO_3$ one should note strong distinction to RS of $LaAlO_3$. The main difference is appearance of new intensive high-frequency line near $640\; cm^{-1}$ and relative increase of intensities of high frequency lines. The appearance of new RS line is clear evidence for lowering symmetry, i.e. that the real space symmetry of rhombohedral $LaMnO_3$ is lower than $R\overline{3}c$. The symmetry lowering could be due to cooperative JT effect. The only JT active mode in $\Gamma$ point of Brillouin zone (BZ) of $R\overline{3}c$ space group is double degenerate $E_g$ mode, which corresponds to the $\{\vec k_{13}\}\tau_5$ mode of ideal perovskite (in Kovalev notation \cite{Kovalev}) or $R_{12}$ (in BSW notation).
%
%
Two components of $E_g$ mode could be indexed as $q_\theta$ and $q_\varepsilon$ (Fig. \ref{FigEg}). 
The symmetry group of new phase depends on the ratio of $q_\theta$ to $q_\varepsilon$. This ratio is convenient to describe by an $\tilde \Phi$ angle, defined as $q_\theta=\tilde \rho \cos \tilde \Phi$ and $q_\varepsilon=\tilde \rho \sin \tilde \Phi$. If $\tilde \Phi=\frac{\pi}{2}, \frac{\pi}{2}\pm\frac{2\pi}{3}$ the structure will be with $P2_1/b$ space symmetry. In other case space symmetry will be $P\overline{1}$. 

The JT distortions in $P2_1/b$ space group differs strongly from that in orthorhombic phase with $Pnma$ space group: in orthorhombic phase the $Q_\varepsilon$ distortions correspond to the $M$ point of ideal perovskite BZ ($\{\vec k_{11} \}\tau_5$ in Kovalev notation \cite{Kovalev}). \textit{ On other hand, in experimentally observed monoclinic phase structure \cite{HuangSantoro} JT distortions correspond to the $R_{12}$ mode of ideal perovskite.}

$P2_1/b$ and $P\overline{1}$ groups both have all 12 gerade modes active in RS.  These modes correspond to the following $R\overline{3}c$ vibrations: split four $E_g$ double degenerate mode; one $A_{1g}$ mode, and three $A_{2g}$ modes.  The symmetry breaking would give rise to the appearance of a new line. It corresponds to the $A_{2g}$ vibration. The intensities calculation revealed the remaining two $A_{2g}$ modes to be low RS active. 
For the investigation of such phase transition we use lattice structure and dynamics calculations.
\section{Lattice Structure and Dynamics Calculations} 
         \subsection{Energy calculation model}
In lattice structure calculation the crystal lattice parameters are obtained by the optimization of crystal energy with respect to displacements of ions that do not change crystal lattice symmetry. The crystal energy is calculated in the framework of ionic pair potential shell model explicitly allowing for the JT term, which is simulated as the sum of lower branches of adiabatic potentials of single $[MnO_6]$ complexes \cite{NikiforovShashkin,NikiforovPopov00}
\begin{equation}\label{EqTotalEnergy}
	E=E_{Lat}+E_{JT}
\end{equation}
In the pairwise-potential approximation and the shell model, the lattice energy can be written as 
\begin{equation}
E_{Lat} = E_{Coul}+\frac{1}{2}\sum_{i}\sum_{k \neq i} {V_{ik}} + \frac{1}{2}\sum_i{k_i \vec \delta_i^2}
\end{equation}
where $E_{Coul}$ describes Coulomb interaction in crystal and the index $i$ specifies ions in the primitive cell, the index $k$ enumerates all ions of the crystal. The term $k_i\delta_i^2$ describes the interaction energy between the core of the ith ion and its shell shifted relative to the core by an amount of $\vec \delta_i$. Short range potential is represented by the Born-Mayer, screening and Van-der-Waals terms
\begin{eqnarray}
	V_{ik}=C_{ik}\exp(-D_{ik}|\vec r_{ik}-\vec \delta_i + \vec \delta_k|) \\ \nonumber
	-\frac{A_{ik}}{r_{ik}}\exp(-B_{ik}r_{ik})-\frac{\lambda_{ik}}{r_{ik}^6}
\end{eqnarray}
The screening term describes the screening of Coulomb potential due to overlap of electronic densities of ion.
Parameters of pair potential model are listed in table \ref{TabPairPotentials}.
The many-particle JT contribution to the crystal energy is approximated by the sum of expressions for the lower branches of adiabatic potentials of the $[MnO_6]$ clusters
\begin{equation}\label{EqJTEnergy}
E_{JT}=-|V_e|\sum_n{\sqrt{Q_\theta^2(n)+Q_\varepsilon^2(n)}}
\end{equation}
where $V_e$ is the linear JT interaction constant ($|V_e|=1.3\; \text{eV/\AA}$ \cite{NikiforovPopov00}), and $Q_\theta(n)$, $Q_\varepsilon(n)$ are projection of displacements of oxygen ions onto \textit{local JT distortions}. Actually, these local distortions are projection of positions of all ions onto $[MnO_6]$ cluster local normal JT active modes. The operation of projection can be performed using Van-Fleck coefficients
\begin{eqnarray} \label{EqVanFleck}
Q_\theta(n)=\sum_{\vec k \gamma }{C_{n\theta,\vec k \gamma}q_{\vec k \gamma}} \\ \nonumber
Q_\varepsilon(n)=\sum_{ \vec k \gamma}{C_{ n\varepsilon,\vec k \gamma } q_{\vec k \gamma}}
\end{eqnarray}
where $n$ - indexes JT clusters, $C_{(\theta,\varepsilon) n \vec k \gamma}$ - is the projection of displacements caused by crystal vibration with wave vector $\vec k$ and number $\gamma$ onto local $Q_{(\theta,\varepsilon)}$ distortion.

         \subsection{Raman scattering spectrum calculation model}

To calculate the phonon frequencies, one should diagonalize the dynamic matrix of the crystal.
\begin{equation}\label{EqDynamicalMatrix}
	D=M\left(F^{CC}-F^{CS}\left[F^{SS}\right]^{-1}F^{SC} + F^{JT} \right)_0
\end{equation} 
where $M_{k\alpha,n\beta}=\delta_{\alpha\beta}[m_k m_n]^{-1/2}$; $F^{CC}$ and $F^{SS}$ are the matrices of the second derivatives of the crystal energy (without the JT contribution) with respect to the displacements of the cores and shells of ions, respectively; $F^{CS}$ and $F^{SC}$ are the matrices of the mixed derivatives of the crystal energy (without the JT contribution) with respect to the displacements of the cores and shells of ions; and $F^{JT}$ is the matrix of the second derivatives of the JT contribution in Eq. \ref{EqJTEnergy} with respect to the ion core displacements. 

The ions shells simulate the polarizability of ions, which arises due to transitions to the excited states with different parity.  The RS process involves virtual transitions to these states. That is why the RS intensities could be estimated in shell model. 
Hence, the intensity can be approximated concerning the crystal like classical media.
\begin{equation}
	I_q \propto \left(\frac{\partial \alpha_{\infty}}{\partial q}\right)^2
\end{equation}
where $I_q$ is the intensity of mode $q$, $\alpha_{\infty}$ is the high frequency polarizability. The derivative is taken with respect to the amplitude of vibration $q$. 
In the shell model approximation the derivative could be calculated as follows:
\begin{eqnarray}
\frac{\partial \alpha_{\infty\, \alpha\beta}}{\partial q_i }=\frac{1}{\Omega} Y_k \left[ U' [F^{SS}]^{-1}F^{SCS}[F^{SS}]^{-1}\right]_{k\alpha,i , m\beta}Y_m \\ \nonumber
U'_{i, n\gamma}=\frac{1}{\sqrt{m_n}}U_{i, n\gamma}
\end{eqnarray}  
where $\Omega$ is the volume of primitive cell, $U$ is the matrix which diagonalizes dynamical matrix of the crystal (Eq. \ref{EqDynamicalMatrix}), $i$ - indexes vibrational modes, $\alpha$ and $\beta$ index Cartesian coordinates, $k$ and $m$ index ions in the primitive cell, $Y_k$ - is the charge of the shell with number $k$, $m_k$ is the mass of the ion with number $k$. $F^{SCS}$ is the matrix of third derivatives taken with respect to the shell ($S$) and core ($C$) displacements. 

Using the same approximation for calculation of frequencies and RS lines intensities we can perform the assignment of RS peaks to the calculated phonons. Moreover, this approximation allows investigating the RS lines intensities dependence on the values model parameters like pair potentials or JT term parameters as well as crystal lattice distortions. 

         \subsection{Results and discussion}

If the JT contribution from lattice energy (Eq. \ref{EqTotalEnergy}) will be removed ($|V_e|=0$ in Eq. (\ref{EqJTEnergy})), the structure, obtained via minimization of energy, would have space group $R\overline{3}c$ (tabl. \ref{TabR3cStructure}). Adding JT term to the crystal energy we will get monoclinic phase with $P2_1/b$ symmetry. Structural phase transition from $R\overline{3}c$ to $P2_1/b$ structure in rhombohedral $LaMnO_3$ was observed \cite{HuangSantoro}. 
We suppose that in the high temperature rhombohedral phase JT effect remains, but it is of cooperative dynamical type. We performed the calculation of energy dependence on the values of $E_g$ vibrational mode components $q_\theta$ and $q_\varepsilon$. For each fixed values of these parameters the crystal energy was optimized with respect to the all lattice parameters which do not breaks new, $P\overline{1}$, symmetry. Because of fixed values of $q_\theta$ and $q_\varepsilon$ parameters there was some constrains on the optimization parameters: a) the projection of ions coordinates onto $q_\theta$, $q_\varepsilon$ displacements were constant; b) the lattice constants were constrained to be like in $R\overline{3}c$ lattice, i.e. the lattice was allowed only to change its volume and elongate along cubic [111] direction (such constrains do not allow for the uniform JT distortion) 

\begin{figure}
\includegraphics[scale=1.5]{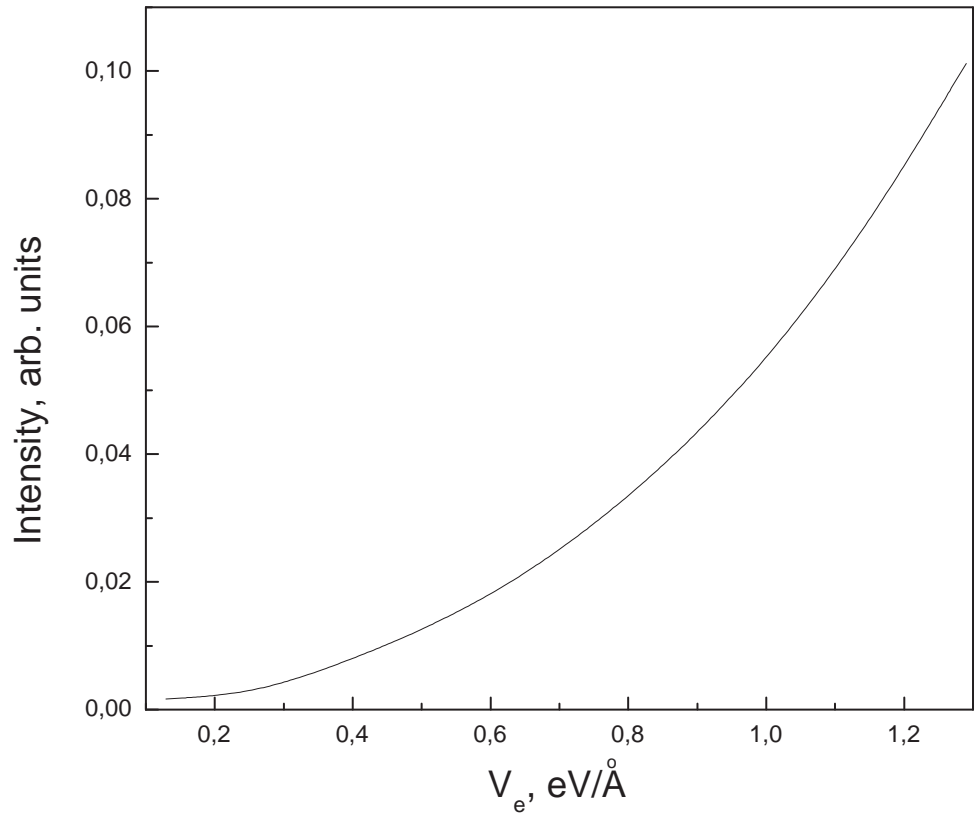}
\caption{\label{FigIntensity} The dependence of intensity of highest $A_{2g}$ (in rhombohedral phase notation) mode on the value of $|V_e|$. The frequency and intensity were averaged on $\tilde \Phi$ for the value of $\tilde \rho$, which corresponds to the energy minimum. For each value of $\tilde \Phi$ the ions were ''relaxed'' according to scheme described in text}
\end{figure}

The calculated dependence could be described well with following relationship (energy per primitive cell)
\begin{equation}
E=\frac{k}{2}\left(q_\theta^2+q_\varepsilon^2\right)-2|V_e|\sqrt{q_\theta^2+q_\varepsilon^2}+
A\left(3 q_\theta^2 q_\varepsilon - q_\varepsilon^3\right)
\end{equation}
Where $k=12\; \text{eV/\AA}^2$, $|V_e|=1.3\; \text{eV/\AA}$ and $A=0.005\; \text{eV/\AA}^3$
Because of small value of $A$ system will perform "slow rotation" at the bottom of adiabatical potential. The trajectory of motion will be approximately described by the equation $\sqrt{(q_\theta^2+q_\varepsilon^2)}=\tilde \rho=2 |V_e|/k$. The value of $\tilde \rho$ is much more grater then the oxygen ions vibrational motions amplitude. We suppose that the frequency of motion at the bottom of adiabatic potential is slower then frequencies of optical phonons. We have to average over the trajectory of "motion" in the $q_\theta$, $q_\varepsilon$ space to get observable physical values. The RS process is "fast" process, but the total time of observation is much grater than process itself. That is why RS process "feels" each low symmetry structure that corresponds to different values of $\tilde \Phi$, but the total picture is averaged on the whole trajectory of motion. In such case we have to average phonon frequencies and their RS intensities over the different $\tilde \Phi$ values. The dynamic matrix is calculated and \textit{diagonalized for each value of $\tilde \Phi$}. 
As the result the averaged phonon frequencies can be got by the averaging over the $\tilde \Phi$ dependence (table \ref{TabRaman}).
The dynamic breaking of symmetry could give rise to the appearance of new RS lines. For verification of such assumption we calculated the dependence of intensity of highest $A_{2g}$ vibration (the symmetry in $R\overline{3}c$\, structure) on the value of $|V_e|$. For each value of $|V_e|$ we calculated the $\tilde \rho$ which corresponds to the energy minimum (the bottom of adiabatic potential). For this value of $\tilde \rho$ the phonon frequencies and RS lines intensities were averaged over all values of $\tilde \Phi$ (Fig.~\ref{FigIntensity}). 

Because of low symmetry at each point of trajectory of averaging the phonon frequencies of $E_g$ symmetry are split (Tab.~\ref{TabRaman}). However, the value of gap is rather small to see distinct lines and to give rise for observed in experiment \cite{Iliev1, Iliev2, MartinCarron02} broadening. The broadening of lines could have the vibronic origin, which is integral feature of systems with dynamical JT effect \cite{BersukerPolinger}.

\begin{acknowledgments}
This study was supported in part by the grant of RFBR R2002 U02-02-96412 and CRDF Grant REC 005.
\end{acknowledgments}

\bibliography{NikiforovPopov}

\newpage
\begin{table*}
\caption{\label{TabPairPotentials} The pair potential model parameters (all in atom units) }  
\begin{ruledtabular}
\begin{tabular}{*{9}{c}}
	&$La^{3+}-O^{2-}$	&$Mn^{3+}-O^{2-}$	&$O^{2-}-O^{2-}$	&	&$La^{3+}$	&$Mn^{3+}$	&$O^{2-}$\\
\hline
A	&85.623	&32.689	&44.862	&X	&9	&8.265	&4.245\\
B	&1.276	&1.671	&1.0592	&Y	&-6	&-6	&-6    \\
C	&250.0	&43.56	&21.729	&k	&40.0	&50.76	&3.99   \\
D	&1.650	&1.575	&1.320	&	&	&	&       \\
$\lambda$	&0	&0	&4.513	&	&	&	&\\	
\end{tabular}
\end{ruledtabular}
\end{table*}

 \begin{table}
 \caption{\label{TabR3cStructure} The structure of $LaMnO_3$ in rhombohedral phase}
 \begin{ruledtabular}
\begin{tabular}{*{9}{c}}
	&$a$	&$c$	&\multicolumn{3}{c}{$O^{2-}$}	\\
	&	&	&x	&y	&z	        \\
\hline
Calc.\footnotemark[1]	&5.809	& 13.894& 0.455&0	&1/4		\\
Aver. Calc\footnotemark[2]	&5.837	& 13.911	& 0.451	&0&1/4	\\
Exp. Ref.~\onlinecite{HuangSantoro}&5.529	&13.335	&0.447	&0&1/4	\\
\end{tabular}
 \end{ruledtabular}
\footnotetext[1]{The crystal lattice parameters were calculated via energy optimization having $|V_e|=0$}
\footnotetext[2]{The crystal lattice parameters were obtained via averaging of optimized structure having $|V_e|=1.29\; \text{eV/\AA}$}
 \end{table}
\begin{table}
 \caption{\label{TabRaman} RS lines frequencies}
 \begin{ruledtabular}
\begin{tabular}{*{6}{c}}
Mode&$LaAlO_3$\footnotemark[1]&$LaMnO_3$\footnotemark[2]&$LaMnO_3$\footnotemark[1]&$LaMnO_3\footnotemark[3]$&$LaMnO_3$\footnotemark[4]\\
\hline
$E_{g} $&33 &39 &-  &-  &61 \\
        &   &   &   &   &74 \\
$A_{2g}$&-  &144&-  &-  &142\\
$E_{g} $&180&172&179&-  &168\\
        &   &   &   &-  &177\\
$A_{1g}$&123&223&236&217&235\\
$A_{2g}$&-  &261&-  &-  &261\\
$E_{g} $&152&312&329&323&314\\
        &   &   &   &   &322\\
$E_{g} $&487&468&515&427&-\footnotemark[5]  \\
        &   &   &   &497&478\\
$A_{2g}$&-  &580&640&618&584\\
\end{tabular}
 \end{ruledtabular}
\footnotetext[1]{The experimental values are taken from Ref.~\onlinecite{Iliev1}. An assignment is carried out according to our intensities calculations.}
\footnotetext[2]{The calculated frequencies without JT term in crystal energy and dynamical matrix ($R\overline{3}c$ space group)}
\footnotetext[3]{The experimental values are taken from Ref.~\onlinecite{Iliev2}}
\footnotetext[4]{An averaged on the AP bottom RS frequencies}
\footnotetext[5]{This mode corresponds to the motion along the bottom of adiabatical potential and to get frequency of it will require separate consideration}
\end{table}
\newpage


\end{document}